# Using Power-Hardware-in-the-Loop Experiments together with Co-simulation for the Holistic Validation of Cyber-Physical Energy Systems

Van Hoa NGUYEN[1], Yvon BESANGER[1], Quoc Tuan TRAN[2], Cédric BOUDINNET[1], Tung Lam NGUYEN[1,2],
Ron BRANDL[3], Thomas I. STRASSER[4]
[1]University Grenoble Alpes, G2Elab, F-38000 Grenoble, France
CNRS, G2Elab, F-38000 Grenoble, France
[2]CEA-INES, 50 Avenue du Lac Léman, 73370 Le Bourget-du-lac, France
[3]Fraunhofer Institute of Wind Energy and Energy System Technology, Kassel, Germany
[4]AIT Austrian Institute of Technology, A-1210 Vienna, Austria
Email: van-hoa.nguyen@grenoble-inp.fr

*Abstract*—Composed of a large variety of technologies and applications with unprecedented complexity, the smart grid, as a cyber-physical energy system, needs careful investigation of the interactions between the various domains involved, especially the coupling between the power and information systems. In this paper, two modern ways of modeling and simulating complex cyber-physical energy systems are considered: Co-simulation and Power-Hardware-in-the-Loop experiments. An analysis of these two approaches shows that a complementary and joint setup with realistic behaviors of hardware equipment under a variety of complex environments, co-simulated by several simulators from different domains, create a complete and high performance environment to achieve a holistic approach for smart grid validation and roll out. In the scope of coupling these two techniques, major technical challenges are identified and advanced solutions are outlined.

## I. INTRODUCTION

A variety of changes and developments have been carried out during the past, presenting a portfolio of initiatives and perspectives of different smart grid solutions on European but also on international level [1]–[3]. In general, an increased consumption of electricity, with important peak loading due to electrification of transport is expected [1]. The decarbonized scenario requires a high penetration of distributed and renewable energy resources, over levels of 15% to 20%, leading to an increasingly difficulty to ensure the reliable and stable management of electricity systems [3]. The integration of Information and Communication Technology (ICT) into the electrical energy infrastructure, along with smart metering is shifting from demonstration phase to large scale deployment. This will have a strong impact on system architectures as well as it raises concerns about cyber-security issue. The electric power grid integrated with communication systems and distributed energy resources (solar, heat, etc.) has become a *cyber-physical energy system (smart grid)* nowadays.

A general framework for smart grid validation and roll out, which takes into account their mutual interactions and interdependencies, is required. One of the main barriers to this has been the lack of design and validation tools that are capable of analyzing power and communication systems in a holistic manner.

Extending power system simulation tools for the ICT domain, or vice versa, demands a lot of effort and collaboration among experts of both areas, because the life cycle and technical specifications of the electrical and communication equipment (in terms of reliability requirements, round trip time, determinism, temporal consistency and hierarchy) are significantly different. By creating a so-called co-simulation environment for the integrated analysis of both domains, via means of ad-hoc connections or in a master/slave fashion, one can understand the impacts of different communication solutions used for the operation of power systems much better. Although simulation architectures may vary, a co-simulation framework allows in general the joint and simultaneous investigation of models developed with different tools, in which the intermediate results are exchanged during the execution of the simulations. However, the sub-systems are usually solved independently by their corresponding domain-specific simulators [4]. Co-simulation allows to have a complete view of both network behavior and the physical energy system states, while power system and communication networks are simulated with the most suitable solver and the calculation loads are shared.

Power-Hardware-in-the-Loop (PHIL) technology is increasingly used by the industry and the research community for testing hardware components, devices or systems in real conditions and scale, where a part of the whole test components are simulated in a Digital Real-Time Simulator (DRTS) [5]. The PHIL approach allows a safe and repeatedly testing of a device also in faulty and extreme conditions without damaging lab equipment, while providing also flexibility in setting up the test setup in transient and steady state operation [5], [6]. In the European ERIGrid project, a survey was addressed to the experts in 12 top European power and energy systems research institutions about mandatory future improvements of PHIL technology[1]. 63.6% of the experts stated that the power and software interfaces in PHIL technology should be im-proved to enable the capacity of remote/distributed testing and integration with co-simulation.

---
[1]81.8% have PHIL testing capacity, in which 54.5% are in well-developed/experienced state.

In this paper, we explore the possibilities of integrating PHIL in co-simulation in order to enable a holistic evaluation of smart grid solutions (addressing mainly power and ICT domains). By offering experiments very closed to real situations, this approach provides an important tool during the design, implementation and roll out of smart grid technologies, solutions and corresponding products. Beyond the added value to testing methods, this approach would also offer international and multi-laboratory cooperation, which in turn, have a positive impact to interoperability and confidence in the applicability of the researches under different grid conditions.

The main parts of the paper are structured as follows: The fundamental points of PHIL and co-simulation relevant for this investigation are outlined in Section II. A general architecture is also proposed in this section. Major technical obstacles towards a seamless integration of PHIL and co-simulation are discussed in Section III together with possible solutions. The paper is concluded with a discussion and an outlook about future research in Section IV.

## II. INTEGRATION OF PHIL AND CO-SIMULATION

In this section, we outline the principal elements of the co-simulation and the PHIL approaches followed by a generic architecture for an integration of them.

### A. Co-simulation of Power and ICT Systems

Most of the work related to co-simulation in the field of smart grid solutions is mainly related to the necessity of interconnecting power system and communication network simulations. As the traditional passive electric power grid (with uni-directional power flows) evolves towards and active power system (with bi-directional power flows), the existing energy infrastructure suffer from several drawbacks (fragmented architecture, lack of adequate bandwidth for two-way communication, in-ability to handle the increasing amount of data from smart devices, etc.) [7]. It is therefore crucial to take the communication network in the development of smart grids – in terms of efficient topology, latency and security – into account.

Usually, communication networks used in the context of lab experiments have very low latency due to short geographical distances. This does not reflect real scenarios whereas the long geographic distance between different, networked devices may cause unexpected delays and signal losses resulting in an unexpected and faulty control behavior. Therefore, the communication network is usually separately analyzed using dedicated software tools in order to study the effect of realistic latencies, packet losses or failures in the ICT/automation system [8]. Communication simulators facilitate also cyber-security related investigations, such as denial-of-service protection, confidentiality and integrity testing.

Co-simulation of the power and the communication system for an integrated analysis of both domains is however, not an easy task since the synchronizing of both simulation packages during runtime is required. Moreover, existing simulation tools are usually provided limited coupling possibilities with external tools; an adequate and suitable Application Programming Interface (API) is often missing. On top of that, the fundamentally different concepts behind power and communication systems are also a challenge; detecting, linking, and handling related events in both domains can be a complex task (cf. Fig. 1)[2]:

- Power system simulation is usually continuous with the possibility of detecting events associated to values crossing a certain threshold.
- Communication network simulation is based on discrete events whose occurrence usually unevenly distributed with respect to time. Corresponding domain-specific simulators provide an event scheduler to record current system time and process the events in an event list.

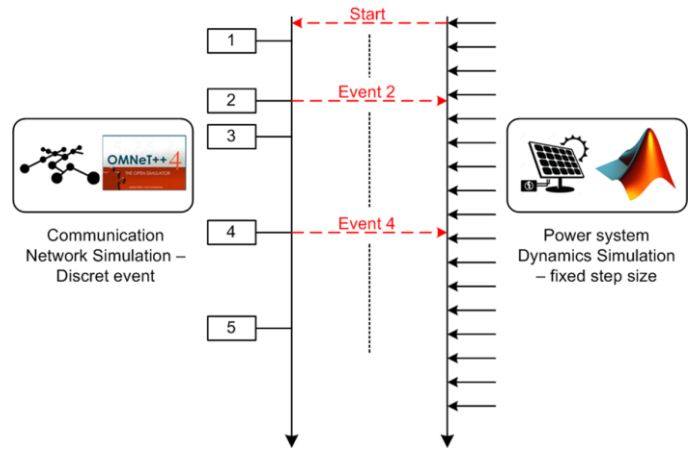

Fig. 1. Time synchronization between power and communication simulation.

Once an event occurs, the associated information is passed to the other domain where the other simulator will create the reaction. A co-simulation framework then has to execute some algorithms to ensure the synchronous and deterministic execution of both domains simultaneously. Scientists have come up with various methods and techniques to deal with the synchronization issue. We can classify them into four main synchronization techniques:

- *"Offline" Co-simulation or "Model Exchange":* In this approach, the model of power system is exported to C-codes and then be compiled and imported to the network simulator for co-simulation. This is usually used as an alternative when direct co-simulation is hard to achieve.
- *Master-Slave:* In this approach, one simulator (usually the communication simulator, due to discrete timeline) is given higher priority and will coordinate the co-simulation steps.
- *Point-based or time-stepped method:* The individual simulators run their simulations independently but pause at fixed synchronization points where information is exchanged between simulators. In this approach, a middleware is normally needed.
- *Global Event-Driven:* In this approach, a global event list is created by mixing up the power system iteration steps with the communication network events according to their timestamps. Then only one simulator is allowed to pro-ceed at a time and the other will halt. This structure leads to a limited speed of co-simulation.

---
[2]The software mentioned in the figures is representative and serves only for illustration purpose, to aide with reader comprehension. They are, by no mean, suggestion or recommendation from the authors for simulator selection.

A quite general review of existed works on the co-simulation of power and communication systems can be found in [7]–[9]. Generally, one can acknowledge two different structures of co-simulation:

- *Ad-hoc Co-simulation:* Most of the work in the literature falls into this category (usually coupling directly one power system simulator and one communication network simulator).
- *Co-simulation with Master Algorithm:* In this approach, a master algorithm (e.g., HLA [10]) or a co-simulation framework (e.g., mosaik[3], Ptolemy[4]) will orchestrate the process. This master algorithm is responsible for synchronizing different timelines of involved simulators and for directing the information exchange among simulator's inputs/outputs.

In order to improve interoperability and reusability of the models developed in co-simulation frameworks mainly two major standards have been issued: *Functional Mockup Interface[5] (FMI)* and *High Level Architecture framework (HLA)* [10]. While FMI is oriented towards model exchange and the coupling of simulators for co-simulation, HLA provides a kind of master algorithm to orchestrate the co-simulated processes (which is addressed as "federate"). It appears that while both standards serve for co-simulation, FMI and HLA are not exactly at the same level of abstraction. Individually, HLA allows highly parallelized simulations of large-scale systems, but introduces additional time-synchronization issues [8].

### B. Power-Hardware-in-the-Loop Experiments

The high ratio of Distributed Energy Resources (DER) integration in a decarbonized scenario leads also to technical difficulty to preserve the security and reliability of the network operation and to ensure the fulfilment of the established voltage quality standards [11]. The Hardware-in-the-Loop (HIL) approach used in the power and energy systems domain is an efficient testing method for DER devices, for manufacturers to adapt their products to the increasingly demanding requirements, as well as for network operators and regulation authorities to establish new testing and certification procedures [6], [11]. In this approach, a real hardware setup for a domain (or part of a domain) is coupled with a simulation tool to allow testing of hardware or software components under realistic conditions. The execution of the simulator in that case requires strictly small simulation time steps in accordance to the real time constraint of the physical target. Since the HIL approach usually involve coupling of different domains, it deals with quite similar challenges as co-simulation. On the other hand, HIL provides the advantage of replacing error-prone or incomplete models with real-world counterparts and the possibility of scalable testing in faulty and extreme conditions.

HIL in smart grids is generally classified into Controller Hardware-in-the-loop (CHIL) and Power-Hardware-in-the-loop (PHIL) experiments [5], [6], [11]. CHIL involves in testing of a device (usually a controller) where signals are exchanged between a DRTS and the device under test via its information ports. The interface consists most of the time only Analogue to Digital and Digital to Analogue converters. In contrary, PHIL involves in testing a device which absorbs or generates power (e.g., inverter-based DER). A power interface is therefore necessary (see Fig. 2). In this paper, the focus is mainly on the PHIL approach.

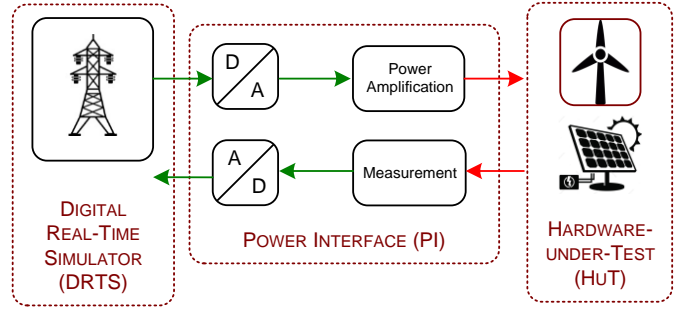

Fig. 2. General architecture of a PHIL experiment.

A general PHIL setup consists of three main elements: *(i)* the DRTS, *(ii)* the Hardware-under-Test (HuT), and *(iii)* the Power Interface (PI):

- The *DRTS* computes the simulation model and offer I/O capacities. As aforementioned, the simulation time-step of the DRTS must be small enough to reproduce the behavior of the simulated system under dynamic condition (Fig. 3). The simulator allows designing and performing various test scenarios with a great flexibility.
- The *HuT* is usually a wide variety of different DER devices and networks (e.g., inverter-based DER, electric vehicles, smart transformers) or a whole microgrid can also be tested in a realistic environment.
- A *PI* generally consists of a power amplifier and sensors that transmit measurements in feedback. It allows the interaction of the virtual simulated system with the HuT.

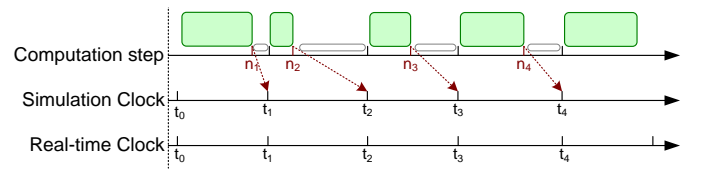

Fig. 3. Time step restriction of a real-time simulation.

While offering a great flexibility of testing, PHIL requires serious consideration on stability and accuracy [11]. The introduction of power interface to the test setup creates an additional close loop, which possibly injects errors, time delay, and distortion that may cause severe instability issues or inaccurate results [12]. Generally, the power amplifier also affects the magnitude and phase of the signal under amplification. However, the inserted time delay is the main obstacle that limits the current capacity of PHIL over a large geographical area or remote PHIL.

The two principal characteristics of a PI in PHIL experiments are the *power amplification unit* and the *interface algorithm*. There is a variety of options for power amplifier with diverse performance characteristics. A review on power amplification units and their topologies can be found in [6]. Comparison of different types of amplifier, as well as recommendations for selection can be found in [13]. In general, the following three types of power amplifiers are common in PHIL experiments:

- *Switched-mode Power Amplification:* Commonly used for small scale PHIL simulation in megawatt range. It is less expensive but represents a higher level of time-delay and lower accuracy than the others.
- *Generator Type Power Amplification:* Is used extensively for interfacing of balanced three phase grid simulations at low and medium power range.
- *Linear Power Amplification:* Is the most suitable aggregate for PHIL applications in small to medium power range. The linear amplifier has very high dynamic performance, short time delay and less stability issues.

---

[3] http://mosaik.offis.de/
[4] http://ptolemy.eecs.berkeley.edu/
[5] https://www.fmi-standard.org/

Configuration and impact of the power amplifier (I/O boundaries, galvanic isolation, short circuit behavior, slew-rate, etc.) must be addressed and evaluated to match the specific requirement for each PHIL setup as it strongly influence the determination of system stability, bandwidth, and the expected accuracy.

The interface algorithm between the DRTS and the hardware part in a PHIL experiment may be either voltage type (for voltage amplifier) or current type (for current amplifier). Three commonly employed interface algorithms are:

- Ideal Transformer Method (ITM)
- Partial circuit duplication method (PCD)
- Damping impedance method (DIM)

A complete review of various interface algorithms and recommendations for selection in PHIL experiments can be found in [11] and [12].

Offering a wide range of possibilities for validation and testing of smart grid solutions, PHIL simulations still be restricted by some limitations, mostly due to the technical challenges related to the introduction of a power interface (e.g., simulation of nonlinearities, studies of high harmonics, stability and power level of amplifier, accuracy of measurements in transient phase, bandwidth limitation). The main difficulty towards integration of PHIL into a holistic validation framework is, inter alia, the issue of signal latency, compensation of loop delay and time synchronization. Besides, the issue of time synchronization also limits the capacity of PHIL simulation of complex systems. Due to the aforementioned obstacles, especially the stability issue, there does not exist any interface or standard which enables interoperable PHIL applications. The harmonization and standardization of PHIL testing is therefore also a topic of common interest to the power and energy domain.

*C. Integration of PHIL and Co-simulation*

We investigate in this section the possibility of integrating PHIL technology in a co-simulation framework in a holistic approach for cyber-physical energy systems. Combining the strong points of both approaches, we can study multi-domain experiments with realistic behaviors from hardware equipment under a variety of complex environments, co-simulated by several simulators from different domains. It will enable a complete consideration of electrical grid interconnected with other domains and is an important contribution to a holistic approach for smart grid system validation and roll out. A general architecture for this integration is proposed in Fig. 4.

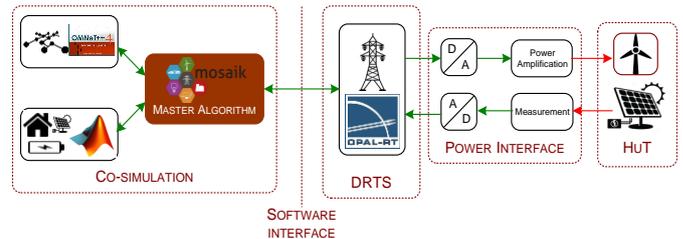

Fig. 4. General architecture for the integration of PHIL and co-simulation.

This architecture also enables the possibility to cooperate multiple research infrastructures' resources to actualize collaborated experiments and provides a way to include valuable knowledge and intelligence of researcher from different domains to study, in a holistic manner, the cyber-physical energy system (i.e., smart grid). This desired scenario requires however a strong interoperability among partner's platforms in various levels.

Most of the current works involving the integration of the HIL approach into a co-simulation framework use only a direct coupling with the DRTS [14] or a kind of CHIL setup [15]. Only until recently, scientists have investigated the possibility of extending PHIL beyond laboratory geographical boundaries, and mostly, for latency tolerant applications, i.e. monitoring [16]. These developments, along with deeper studies on impact of latency in distributed DRTS [17], would create a technical base to enable the integration of PHIL to co-simulation framework.

III. TECHNICAL CHALLENGES AND PROPOSED SOLUTIONS

In order to run the holistic experiment correctly and seamlessly, the following major technical challenges arise.

*A. Data Flow and Concurrency*

Within the process of integrating PHIL into co-simulation, it is crucial to ensure a synchronous data flow among the individual components, as well as the concurrency of the different simulators. In the general architecture from Fig. 4, three spots should be considered:

*1) Power Interface*

Basically, the challenge here is to synchronize and compensate the loop delay in order to stabilize the system and increase the accuracy of the test. The first step should be the selection of an appropriated interface algorithms and corresponding power amplification where recommendations from [11] and [12] should be considered.

Secondly, a time delay compensation method could be applied, such as introducing phase shifting, low-pass filter to the feedback signal [18], extrapolation prediction to compensate for time delays [19], phase advance calibration [20] or multi-rate real-time simulation [21].

*2) Co-simulation Interface*

The issue at the co-simulation interface was already addressed in Section II.A. Besides synchronizing time steps of different simulators using the aforementioned techniques, the master algorithm or the co-simulation framework has to deal with the harmonization of the continuous/discrete event timelines of the power/communication interface.

*3) Software Interface*

On top of that, when integrating with real-time simulation and PHIL, it is necessary to ensure that the harmonized time steps should be small enough to be coupled in real-time. Therefore, the interfacing of real-time and offline simulations needs to be taken into consideration.

The principle of real-time simulation is presented in Fig. 3. Offline simulation, on the other hand, may have a simulation clock speed different to real time clock. Two kinds of non-real-time simulations are classified *(i)* slow, and *(ii)* fast as depicted in Fig. 5.

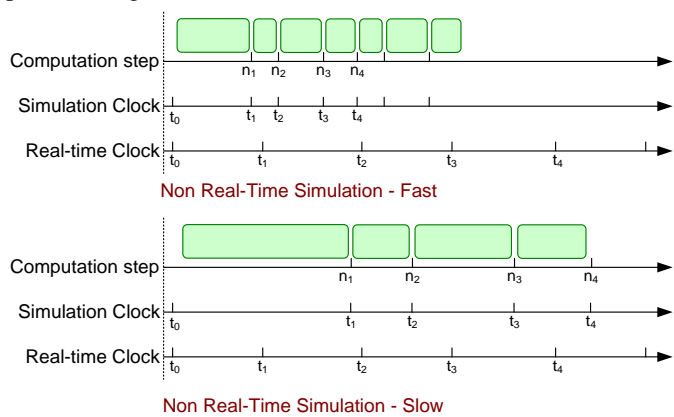

Fig. 5. Non-real-time simulation types.

The non real-time simulations step, in case of coupling, has to be adapted to the real-time simulation step, either by delaying the step in case of fast simulation, or increasing the computation speed in case of slow simulation.

*B. Interoperability, Data Model, and System Topology*

Besides the above technical challenges, when the experiments involve multiple domains or multi-laboratory, it is required to have a certain degree of interoperability among the different actors as well as among different elements of the experiments. A common information model or at least a conversion interface is necessary.

In a power system simulation, the exact and proper representation of a system's topology is critical, proportionally with scale and complexity. The information model should be capable to represent, encapsulate and exchange static and dynamic data, as well as, to inform any modification in topology and current state of the network in real-time and in a standardized way. It is suggested in [22] that IEC 61970/61968 (CIM/XML/RDF) and OPC UA could be combined to provide a seamless and meaningful communication among applications and a strong support for multiplatform experiment, which is capable of transmitting static and dynamic data of system's topology in real-time. This combination, however, does not cover the ICT domain, an interface with the communication simulators must be provided as well.

*C. Remote coupling PHIL/Cosimulation*

In the context of coupling PHIL and co-simulation for multi-laboratory experiments, there are scenarios where the DRTS and offline simulation are geographically separated. In that case, the latency may accumulate and surpass the limitation of the time synchronization algorithm (see Fig. 6). Moreover, random packet loss due to network congestion outside of a communication network (e.g., LAN) may alter the information and cause malfunction to the DRTS, as well as any connected hardware.

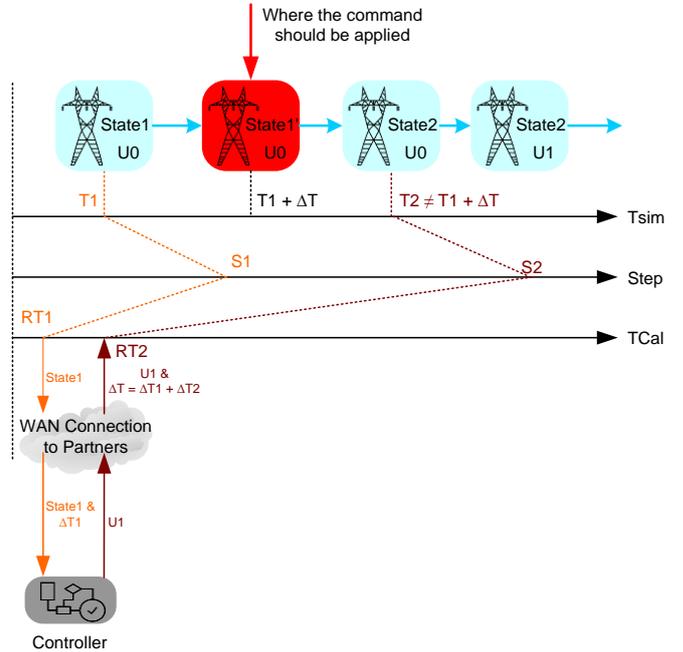

Fig. 6. Delayed application of command due to unexpected latency.

Therefore, in case of coupling PHIL together with co-simulation in geographically distributed experiments, the time synchronization algorithm must be adapted. The PI compensation has to take into account communication latency.

IV. DISCUSSION AND OUTLOOK

Two modern ways of modeling and simulating complex cyber-physical energy systems were presented. While co-simulation includes and combines knowledges in various domains in order to consider the system in a holistic manner, PHIL provides users with the advantage of replacing error-prone or incomplete models with real-world counterparts and the possibility of scalable testing in faulty and extreme conditions. An analysis of these two tools shows that it will make sense to combine the strong points of both approaches to study multi-domain experiments. The advantage is a complementary and joint setup, benefited from realistic behaviors of hardware equipment under a variety of complex environments, co-simulated by several simulators from different domains. The goal is to create a complete and high performance environment to achieve a holistic approach for smart grid validation and roll out.

Major technical challenges have been identified and some solutions were suggested. This contribution gives way for further proposals in future developments of coupling PHIL and co-simulation.

ACKNOWLEDGEMENT

This work is supported by the European Community's Horizon 2020 Programme (H2020/2014-2020) under project "ERIGrid" (Grant Agreement No. 654113, www.erigrid.eu).

The work of G2Elab and CEA-INES is also partially supported by the Carnot Institute "Energies du Futur" under the PPInterop II project (www.energiesdufutur.eu).